\documentclass[aps,prd,twocolumn, preprintnumbers, floatfix, superscriptaddress, nofootinbib]{revtex4}
\usepackage[dvips]{graphics}
\usepackage{amsmath,amsfonts,bm}
\usepackage{epsfig,bm}
\usepackage{graphicx}
\def\MeijG[#1][#2][#3][#4][#5][#6]{G^{#1}_{#2}\left(#3,#4\left|\begin{array}{c}#5\\#6\end{array}\right|\right)}

\begin{document}

\preprint{LMU-ASC~08/14}
\preprint{FLAVOUR(267104)-ERC-63}

\affiliation{Ludwig-Maximilians-Universit\"at M\"unchen, Fakult\"at f\"ur Physik,\\
Arnold Sommerfeld Center for Theoretical Physics, 
80333 M\"unchen, Germany}
\affiliation{TUM-IAS, Lichtenbergstr. 2a, D-85748 Garching, Germany}
\affiliation{Physik Department, TUM, D-85748 Garching, Germany}

\author{Oscar Cat\`a}
\affiliation{Ludwig-Maximilians-Universit\"at M\"unchen, Fakult\"at f\"ur Physik,\\
Arnold Sommerfeld Center for Theoretical Physics, 
80333 M\"unchen, Germany}
\affiliation{TUM-IAS, Lichtenbergstr. 2a, D-85748 Garching, Germany}
\affiliation{Physik Department, TUM, D-85748 Garching, Germany}

\title{\large{Lurking pseudovectors below the TeV scale}}

\begin{abstract}
If electroweak symmetry breaking is driven by a new strongly-coupled dynamical sector, one expects their bound states to appear at the TeV scale or slightly below. However, electroweak precision data imposes severe constraints on most of the existing models, putting them under strain. Conventional models require the new composite states to come in pairs of rather heavy, close to degenerate spin-1 resonances. In this paper I argue that spin-1 states can actually be lighter without clashing with experimental bounds. As an example, I consider a composite model with a light pseudovector resonance that couples to the Standard Model gauge boson, fermion and scalar fields. I show how such a resonance leaves basically no imprint on the NLO corrections to the Standard Model. This happens not through parameter tuning, but rather as a consequence of generic properties of realistic UV completions. This pseudovector is mostly unconstrained by existing data and could be as light as $600$ GeV. In the last part of the paper I briefly discuss its most characteristic signatures for direct detection at colliders.     
\end{abstract}

\maketitle


\section{Introduction}\label{sec:I}

The recent discovery of a scalar sector in the Standard Model has profound implications for particle physics, yet it does not settle the issue of what dynamics is behind electroweak symmetry breaking and how the hierarchy problem is resolved. In that respect, the discovery of new particles in the sub-TeV region, either from weakly-coupled or strongly-coupled dynamical extensions, should provide valuable hints. 

The absence (so far) of new physics states, together with the Higgs-like character of the $126$ GeV scalar, indicates that deviations from the Standard Model paradigm have to be small. This, together with flavor constraints, puts under strain, if it does not already exclude, most of the natural realizations of Supersymmetry and composite models.

In this paper I will concentrate on scenarios of strongly-coupled dynamics at the TeV scale. Composite models were initially introduced as higgsless alternatives to the Standard Model, borrowing heavily from patterns and characteristics of QCD. While those models are nowadays ruled out, they have been superseded by more elaborate strongly-coupled scenarios that accomodate a light scalar through the vacuum misalignment mechanism discussed in~\cite{Kaplan:1983fs,Dugan:1984hq}. In those dynamical scenarios the Higgs-like particle is a pseudo-Goldstone mode of some broken global symmetry which develops a potential through quantum corrections, thereby avoiding the hierarchy problem. One of the virtues of the vacuum misalignment is that it is flexible enough to smoothly interpolate between non-decoupling and decoupling new-physics scenarios. Recently, such ideas have been formulated in an effective field theory language~\cite{Feruglio:1992wf,Bagger:1993zf,Azatov:2012bz,Buchalla:2012qq,Alonso:2012px,Buchalla:2013rka,Buchalla:2013eza}, which allows for a systematic scrutiny of deviations from the Standard Model paradigm. 

While an effective field theory is a general description of the physics at low energies, specific models are needed to identify potential signatures and guide searches at colliders. However, without imposing rather {\it{ad hoc}} conditions, the presently allowed size of new physics effects seems hard to accommodate within models of light (sub-TeV) states. For instance, constraints on oblique parameters typically push the mass-range of vector and axial resonances at a few TeV~\cite{Pich:2012dv,Pich:2013fea}, or else one is lead to a certain degree of fine-tuning by requiring unnaturally sizable one-loop corrections~\cite{Barbieri:2008cc,Cata:2010bv}. 

In this paper I will assume that no large one-loop effects are present, which otherwise would jeopardize the effective field theory expansion, and therefore accept that composite models seem to prefer rather heavy vector and axial states. The question I will then address is whether other lighter spin-1 states can be present. Such states should leave a rather subtle phenomenological imprint in order not to upset electroweak precision measurements but might leave at the same time rather clean signals for direct detection. In the following I will work with a minimal dynamical setting, assuming that the new dynamics is invariant under $SU(2)_L\times SU(2)_R$ broken down to the custodial $SU(2)_V$. This coset structure guarantees that the number of pseudo-Goldstone bosons does not exceed the experimentally-observed ones. In order to be fully general, a light Higgs-light scalar will be introduced as a singlet. I will further assume that this new dynamics is parity-preserving and generates its lightest resonance state roughly around $700-900$ GeV. This state will be described by an antisymmetric rank-two tensor field ${\cal{B}}_{\mu\nu}$ with the quantum numbers of a pseudovector. 

This new dynamics will be assumed to be dual to a more fundamental theory of constituents, which sets in at sufficiently high energies. I will show that consistency with this dual picture implies that a pseudovector would leave no trace on effective operators with gauge bosons. At energies much lower than its mass, its effects would only be noticeable as anomalous top quark vertices. Its impact on low-energy effective operators is thus rather elusive, but still it provides distinct signatures for direct detection at colliders, mainly through gluon-fusion ($gg\to {\cal{B}}\to W^+W^-$) and associated production ($pp\to {\cal{B}} Z$).    

This paper is organized as follows: in Section~\ref{sec:II}, I discuss some technical aspects of rank-two antisymmetric tensors, such as its decomposition in longitudinal and transverse modes and their connection through a duality transformation. In Section~\ref{sec:III} the model for composite pseudovectors is introduced. Its indirect traces at low energies are examined in Section~\ref{sec:IV} while comments on the most promising signatures for direct detection are addressed in Section~\ref{sec:V}. Concluding remarks are given in Section~\ref{sec:VI}. 


\section{Kalb-Ramond fields and massive spin-1 states}\label{sec:II}

It is well-known that massive spin-1 states can be described by one-form Proca fields as well as two-form fields. The Proca description successfully accounts for the Standard Model fundamental gauge spin-1 fields ($W^{\pm},Z$), whose masses are generated after gauge symmetry is spontaneously broken via the Higgs mechanism. In contrast, the two-form representation seems to be better suited for composite spin-1 states. 

In QCD, for instance, the coefficients of the NLO low-energy expansion (the Gasser-Leutwyler coefficients) are expected to be ${\cal{O}}(f^2/\Lambda^2)$, with $f$ the pion decay constant and $\Lambda$ the scale of hadronic physics. This is the natural size one would expect from integrated-out hadronic degrees of freedom lying at $m\sim {\cal{O}}(\Lambda)$ and agrees rather well with experimental data. A two-form representation of axial and vector mesons naturally accounts for this pattern of vector meson exchange~\cite{Ecker:1988te} besides reproducing other key aspects of low-energy chiral dynamics~\cite{Gasser:1983yg,Ecker:1989yg}. If mesons are represented by Proca fields, however, the contribution to the NLO coefficients from tree level resonance exchange vanishes altogether. 

Following the example of QCD, I will henceforth use the tensorial representation. In this Section I will lay out some formal aspects of massive rank-two antisymmetric tensors that will be useful for the Sections to come. I will pay special attention to the interplay of the longitudinal and transverse components of the two forms and a duality transformation connecting them.

Let us start by examining the kinetic term. The most general quadratic form for a second rank antisymmetric tensor with well-defined parity is given by
\begin{align}
{\cal{L}}_H&=\frac{a}{2} \partial_{\sigma}H_{\mu\nu}\partial^{\sigma}H^{\mu\nu}+b\, \partial^{\mu}H_{\mu\nu}\partial_{\lambda}H^{\lambda\nu}+\frac{c}{2} H_{\mu\nu}H^{\mu\nu}
\end{align}
Generically, the structure of the previous Lagrangian does not furnish a representation of the Lorentz group. The generic propagator contains two potential poles:
\begin{equation}\label{proj}
\Delta_{\mu\nu;\lambda\rho}=\frac{P^{T}_{\mu\nu;\lambda\rho}}{aq^2+c}+\frac{P^{L}_{\mu\nu;\lambda\rho}}{(a+b)q^2+c}
\end{equation}
where
\begin{align}\label{projectors}
P^{\mu\nu;\lambda\rho}_L&=g^{\mu\lambda}\frac{q^{\nu}q^{\rho}}{2q^2}-g^{\mu\rho}\frac{q^{\nu}q^{\lambda}}{2q^2}-g^{\nu\lambda}\frac{q^{\mu}q^{\rho}}{2q^2}+g^{\nu\rho}\frac{q^{\mu}q^{\lambda}}{2q^2}\nonumber\\
P^{\mu\nu;\lambda\rho}_T&=-\epsilon^{\mu\nu\alpha\beta}\epsilon^{\lambda\rho\eta\sigma}g_{\alpha\eta}\frac{q_{\beta}q_{\sigma}}{2q^2}
\end{align}
are the transverse and longitudinal projectors for rank-two tensors. As such, $P_{T}^{\mu\nu;\lambda\rho}+P_{L}^{\mu\nu;\lambda\rho}=I^{\mu\nu;\lambda\rho}$, with $I^{\mu\nu;\lambda\rho}=\frac{1}{2}(g^{\mu\lambda}g^{\nu\rho}-g^{\mu\rho}g^{\nu\lambda})$, as can be easily checked from~(\ref{projectors}).

The longitudinal mode is isolated by picking $a=0$, $b=-\frac{1}{2}$ and $c=\frac{m_L^2}{2}$, and leads to
\begin{align}\label{rho}
\Delta_{\mu\nu;\lambda\rho}^L&=\frac{2}{q^2-m_L^2}\left[\frac{q^2-m_L^2}{m_L^2}I_{\mu\nu;\lambda\rho}-\frac{q^2}{m_L^2}P^L_{\mu\nu;\lambda\rho}\right]
\end{align}
which corresponds to the Lagrangian
\begin{align}
{\cal{L}}_L&={\mathrm{tr}}\left[-\partial^{\mu}{\cal{V}}_{\mu\nu}\partial_{\rho}{\cal{V}}^{\rho\nu}+\frac{m_{\cal{V}}^2}{2}{\cal{V}}_{\mu\nu}{\cal{V}}^{\mu\nu}\right]
\end{align}
where the trace is over the internal symmetry group. The longitudinal field ${\cal{V}}_{\mu\nu}$ is the one currently used to represent spin-1 mesons in QCD~\cite{Ecker:1988te,Gasser:1983yg}. 

The transverse mode corresponds instead to the parameter choice $a=-b=\frac{1}{2}$ and $c=-\frac{m_T^2}{2}$, giving
\begin{align}
\Delta_{\mu\nu;\lambda\rho}^T&=- \frac{2}{q^2-m_T^2}\left[\frac{q^2-m_T^2}{m_T^2}I_{\mu\nu;\lambda\rho}-\frac{q^2}{m_T^2}P^{T}_{\mu\nu;\lambda\rho}\right]\nonumber\\
&=\frac{2}{q^2-m_T^2}\left[I_{\mu\nu;\lambda\rho}-\frac{q^2}{m_T^2}P^L_{\mu\nu;\lambda\rho}\right]
\end{align}
which results from the Lagrangian
\begin{align}\label{KR}
{\cal{L}}_T&={\mathrm{tr}}\left[\frac{1}{2}\partial_{\lambda}{\cal{B}}^{\mu\nu}\partial^{\lambda}{\cal{B}}_{\mu\nu}-\partial^{\mu}{\cal{B}}_{\mu\nu}\partial_{\rho}{\cal{B}}^{\rho\nu}-\frac{m_{\cal{B}}^2}{2}{\cal{B}}_{\mu\nu}{\cal{B}}^{\mu\nu}\right]\nonumber\\
\end{align}
The transverse mode is the natural extension of the gauge two-form field when a mass term is added. A gauged two form is defined by the free action
\begin{align}\label{gaugeH}
{\cal{L}}&=\frac{1}{6}{\mathrm{tr}}\left[H_{\mu\nu\rho}H^{\mu\nu\rho}\right]
\end{align}
where $H_{\mu\nu\lambda}\equiv \partial_{\mu}H_{\nu\lambda}+\partial_{\nu}H_{\lambda\mu}+\partial_{\lambda}H_{\mu\nu}$ is the curvature tensor and the metric signature is chosen mostly negative. The previous Lagrangian is invariant under the gauge symmetry $\delta H_{\mu\nu}=\partial_{\mu}\Lambda_{\nu}-\partial_{\nu}\Lambda_{\mu}$, which eventually leaves only one independent degree of freedom. It therefore describes a massless spin-0 mode, as can be seen by direct investigation of its helicity structure~\cite{Ogievetsky:1967ij} or through duality~\cite{Cremmer:1973mg}.

In the context of string theory, the field $H_{\mu\nu}$ (both massless and massive) was found to be the natural dynamical object mediating interstring interactions~\cite{Cremmer:1973mg,Kalb:1974yc} and is commonly referred to as the Kalb-Ramond field $B_{\mu\nu}$. There, the gauged two form gets its mass through mixing with a gauge one-form field, absorbs its two degrees of freedom and eventually describes a massive spin-1 mode. The opposite also holds true, namely a gauge field can be made massive by absorbing the scalar degree of freedom hidden inside the gauged two form~\cite{Allen:1990gb}.

Eq.~(\ref{KR}) can be easily cast as
\begin{align}
{\cal{L}}_T&={\mathrm{tr}}\left[\frac{1}{6}{\cal{B}}_{\mu\nu\lambda}{\cal{B}}^{\mu\nu\lambda}-\frac{m_{\cal{B}}^2}{2}{\cal{B}}_{\mu\nu}{\cal{B}}^{\mu\nu}\right]
\end{align}
which shows that ${\cal{B}}_{\mu\nu}$ is a massive Kalb-Ramond field.

Therefore, in full generality, a second rank antisymmetric tensor $H_{\mu\nu}$ has 6 degrees of freedom, which can be decomposed as two massive spin-1 fields, the transverse (Kalb-Ramond) ${\cal{B}}_{\mu\nu}$ and the longitudinal ${\cal{V}}_{\mu\nu}$. 

\subsection{Duality transformation}\label{subsec:II.I}

The existence of two tensorial representations for massive spin-1 fields can also be understood by the fact that there are two independent tensor structures for 1-particle creation matrix elements, namely 
\begin{align}
\langle 0| {\cal{V}}^X_{\mu\nu}|X\rangle&\propto\frac{i}{m_X}(p_{\mu}\epsilon_{\nu}-p_{\nu}\epsilon_{\mu})
\end{align}
and
\begin{align}
\langle 0| {\cal{B}}^X_{\mu\nu}|X\rangle&\propto\frac{i}{m_X}\varepsilon_{\mu\nu\lambda\rho}\epsilon^{\lambda}p^{\rho}
\end{align}
which are the normalizations leading to the propagators discussed above. The previous equations suggest that there is a duality transformation between longitudinal and transverse fields given by
\begin{align}\label{third}
{\cal{V}}_{\mu\nu}\to \frac{1}{2}\varepsilon_{\mu\nu\lambda\rho}{\cal{B}}^{\lambda\rho}
\end{align}
To be more precise, one can show that a theory generically given by
\begin{align}
{\cal{L}}&={\mathrm{tr}}\left[-\partial^{\mu}{\cal{V}}_{\mu\nu}\partial_{\rho}{\cal{V}}^{\rho\nu}+\frac{m^2}{2}{\cal{V}}_{\mu\nu}{\cal{V}}^{\mu\nu}+{\cal{V}}_{\mu\nu}J_{\cal{V}}^{\mu\nu}\right]
\end{align}
where the interactions are built to describe a particle species $X$, is dual to another theory 
\begin{align}
{\cal{L}}&={\mathrm{tr}}\left[\frac{1}{6}{\cal{B}}_{\mu\nu\lambda}{\cal{B}}^{\mu\nu\lambda}-\frac{m^2}{2}{\cal{B}}_{\mu\nu}{\cal{B}}^{\mu\nu}+{\cal{B}}_{\mu\nu}J_{\cal{B}}^{\mu\nu}\right]
\end{align}
which describes the same particles $X$ provided that $J_{\cal{B}}^{\mu\nu}=\frac{1}{2}\varepsilon^{\mu\nu\lambda\rho}J_{{\cal{V}}\lambda\rho}$. As a corollary, in the absence of masses, the original gauge transformation of ${\cal{B}}_{\mu\nu}$, $\delta {\cal{B}}_{\mu\nu}=\partial_{\mu}\Lambda_{\nu}-\partial_{\nu}\Lambda_{\mu}$ is cast in terms of ${\cal{V}}_{\mu\nu}$ as $\delta {\cal{V}}_{\mu\nu}=\epsilon_{\mu\nu\lambda\rho}\partial^{\lambda}\Lambda^{\rho}$. 

The duality above means that every spin-1 particle can be equivalently expressed in terms of either a longitudinal or a transverse two-form field. In theories with parity conservation, particles with the same charge conjugation but opposite parity can be described with ${\cal{V}}_{\mu\nu}$ and ${\cal{B}}_{\mu\nu}$ fields, respectively, while leaving the form of the interaction terms untouched. Thus, if an interacting theory for a regular $1^{--}$ vector is described by a longitudinal field ${\cal{V}}_{\mu\nu}$, the corresponding theory for a pseudovector $1^{+-}$ can be constructed simply by replacing ${\cal{V}}_{\mu\nu}\to{\cal{B}}_{\mu\nu}$ in the interaction terms. This observation will be used in the following Section.


\section{A model for composite pseudovectors}\label{sec:III}

If new strongly-coupled dynamics trigger electroweak symmetry breaking, a mass gap will typically be generated between their associated Goldstone bosons (at the electroweak scale $v$) and bound states (typically at the TeV scale, $\Lambda\sim 4\pi v$ or slightly below). In order to work with a minimal model, the following extra assumptions will be made: (i) the new strong sector is parity-conserving and invariant under $SU(2)_L\times SU(2)_R$, spontaneously broken down to the diagonal subgroup $SU(2)_V$. This is the minimal coset structure required to give gauge bosons a mass; (ii) the Higgs-like scalar originates from spontaneous breaking of an extended space-time or internal global symmetry, which occurs at a higher energy scale. I will not model such breaking and instead I will generically introduce the Higgs as a singlet under the Standard Model gauge group; (iii) the first state in the resonance spectrum is a light $J^{PC}=1^{+-}$ pseudovector mode in the mass-range $600-1000$ GeV. Vector $1^{--}$ and axial $1^{++}$ excitations are above the TeV, as suggested by electroweak precision constraints.    

Pseudovector states are not as exotic as it might seem at first. In QCD their lowest-lying candidate is the $b_1(1235)$ meson, which is slightly lighter than the first axial excitation, $a_1(1260)$. Their properties and phenomenology have been explored in a number of papers, from dispersive analysis involving spectral sum rules~\cite{Craigie:1981jx,Cata:2008zc,Cata:2009fd} to holographic models for spin-1 states~\cite{Cappiello:2010tu,Domokos:2011dn,Alvares:2011wb,Domokos:2012da}, including its low-energy impact on chiral couplings~\cite{Ecker:2007us}. 

Here I will be assuming that, unlike in QCD, the pseudovector resonance is the lowest-lying state of the composite spectrum. It will be described by a Kalb-Ramond field ${\cal{B}}_{\mu\nu}$, transforming as a triplet under the unbroken custodial group:
\begin{align}
{\cal{B}}_{\mu\nu}\to \xi_V{\cal{B}}_{\mu\nu}\xi_V^{\dagger}
\end{align}
where $\xi_V\in SU(2)_V$.  

The Goldstone bosons coming from spontaneous $SU(2)_L\times SU(2)_R\to SU(2)_V$ breaking will be collected in a nonlinear matrix $U$, which transforms as a bifundamental:
\begin{align}
U\to \xi_L U \xi_R^{\dagger}  
\end{align}
where $\xi_{L,R}\in SU(2)_{L,R}$. A convenient parametrization is
\begin{align}
U={\mathrm{exp}}(2i\Phi/v),\quad \Phi=\phi^at^a=\frac{1}{\sqrt{2}}\left(
\begin{array}{cc}
\frac{\phi^0}{\sqrt{2}}& \phi^+ \\
\phi^- & -\frac{\phi^0}{\sqrt{2}} 
\end{array}\right)
\end{align}
Since only $SU(2)_L\times U(1)_Y$ is gauged, its covariant derivative is given by
\begin{align}
D_{\mu}U=\big[\partial_{\mu}+igW_{\mu}-ig^{\prime}B_{\mu}\tau_L\big]U
\end{align}
where $\tau_L=Ut_3U^{\dagger}$. In practice it will be useful to also define $L_{\mu}=iUD_{\mu}U^{\dagger}$.

In order to couple the pseudovector to the Standard Model fields in a gauge-invariant way, it is convenient to introduce quantities that transform only under $SU(2)_V$. This can be done with the aid of the more fundamental field $u$
\begin{align}
u\to \xi_Lu\xi_V^{\dagger}=\xi_Vu\xi_R^{\dagger},\quad u^2=U
\end{align}
The building blocks covariant under $SU(2)_V$ are then
\begin{align}
u^{\dagger}\psi_L&\to \xi_V u^{\dagger}\psi_L\nonumber\\
u\psi_R&\to \xi_V u\psi_R\nonumber\\
u_{\mu}&\to \xi_V u_{\mu}\xi_V^{\dagger}\nonumber\\
f_{+}^{\mu\nu}&\to \xi_V f_{+}^{\mu\nu}\xi_V^{\dagger}
\end{align}
where $u_{\mu}\!=-u^{\dagger}L_{\mu}u$ and $f_{+}^{\mu\nu}\!=u^{\dagger}gW^{\mu\nu}u+ug^{\prime}B^{\mu\nu}t_3u^{\dagger}$. Further derivatives on the fields can be shown to be redundant.

The covariant derivative on ${\cal{B}}_{\mu\nu}$ is defined as 
\begin{align}
\nabla_{\lambda}{\cal{B}}_{\mu\nu}&=\partial_{\lambda}{\cal{B}}_{\mu\nu}+[\Gamma_{\lambda},{\cal{B}}_{\mu\nu}]
\end{align}
with
\begin{align}
\Gamma_{\lambda}&=\frac{1}{2}\bigg[u(\partial_{\lambda}-ig^{\prime}B_{\lambda}t_3)u^{\dagger}+u^{\dagger}(\partial_{\lambda}-igW_{\lambda})u\bigg]
\end{align}
 
The model of electroweak interactions at energies above $m_{\cal{B}}$ will be written as
\begin{align}\label{model}
{\cal{L}}&={\cal{L}}_K+{\cal{L}}_M+{\cal{L}}_{\cal{B}}
\end{align}  
where
\begin{align}
{\cal{L}}_K&=-\sum_{X}\frac{1}{4}X_{\mu\nu}^aX^{\mu\nu\,a}+i\sum_j{\bar{\psi}}_j\gamma_{\mu}D^{\mu}\psi_j+\frac{1}{2}\partial_{\mu}h\partial^{\mu}h\nonumber\\
{\cal{L}}_M&\!=\!\frac{v^2}{4}{\mathrm{tr}}[L_{\mu}L^{\mu}]\zeta(h)\!-\!v\big({\bar{\psi}}_iY_{ij}(h)UP_{\pm}\psi_j\!+{\mathrm{h.c.}}\big)\!-V(h)
\end{align}
is the leading order electroweak Lagrangian in the notation of~\cite{Buchalla:2013rka}, where $X$ is a generic gauge boson and $h$ the Higgs-like scalar. $P_{\pm}$ simply project out the first and second element of the doublet $\psi_R$, respectively. Since $h$ is introduced as a singlet, $V(h), \zeta(h)$ and $Y_{ij}(h)$ are generic functions of $h$, whose particular coefficients are model-dependent. 

The pseudovector sector is chosen as 
\begin{align}\label{modelB}
{\cal{L}}_{\cal{B}}&={\mathrm{tr}}\left[\frac{1}{2}\nabla_{\lambda}{\cal{B}}^{\mu\nu}\nabla^{\lambda}{\cal{B}}_{\mu\nu}-\nabla^{\mu}{\cal{B}}_{\mu\nu}\nabla_{\rho}{\cal{B}}^{\rho\nu}-\frac{m_{\cal{B}}^2}{2}{\cal{B}}_{\mu\nu}{\cal{B}}^{\mu\nu}\right]\nonumber\\
&\!\!\!\!\!\!\!\!+h_t({\bar{\psi}}_L\sigma^{\mu\nu}\!u{\cal{B}}_{\mu\nu}uP_+\psi_R)\!+h_b({\bar{\psi}}_L\sigma^{\mu\nu}\!u{\cal{B}}_{\mu\nu}uP_-\psi_R)\!+{\mathrm{h.c.}}\nonumber\\
&\!\!\!\!\!\!\!\!+{\mathrm{tr}}\bigg[f_1{\cal{B}}_{\mu\nu}f_+^{\mu\nu}+if_2{\cal{B}}_{\mu\nu}[L^{\mu},L^{\nu}]\bigg]
\end{align}
which contains the most general interaction terms linear in ${\cal{B}}$. Interactions with $h$ can be incorporated by appending arbitrary functions of $h$ to those operators. Here I will leave such functions implicit. 

Based on naive dimensional counting, one expects the interaction coefficients to gauge bosons to be $f_i\sim {\cal{O}}(v)$. Regarding the fermion couplings, chirality suggests that $h_j$ are proportional to the Yukawa couplings. Accordingly, the pseudovector couples mostly to the third family of quarks. In the following, $\psi$ will stand for a $(t,b)$ $SU(2)$ doublet. 


\section{Indirect signals}\label{sec:IV}
The presence of a pseudovector in the sub-TeV region leaves potential imprints on the low-energy theory through contributions to anomalous couplings. The size of these contributions will show up as ${\cal{O}}(v^2/m_{\cal{B}}^2)$ NLO corrections and can be captured by an effective field theory (EFT) analysis provided that $m_{\cal{B}}$ is not too light. To be more quantitative, since the NLO coefficients in the EFT are expected to be ${\cal{O}}(v^2/\Lambda^2)$, we require $m_{\cal{B}}> \frac{\Lambda}{5}\sim 600$ GeV in order not to upset the naive power counting. 

Indirect pseudovector effects can be seen by integrating it out from the model of the previous Section. The resulting effective Lagrangian will match onto a subset of the NLO terms of the EFT worked out in~\cite{Buchalla:2013rka}. Writing the model as
\begin{align}
{\cal{L}}&={\mathrm{tr}}\left[\frac{1}{6}{\cal{B}}_{\mu\nu\lambda}{\cal{B}}^{\mu\nu\lambda}-\frac{m_{\cal{B}}^2}{2}{\cal{B}}_{\mu\nu}{\cal{B}}^{\mu\nu}+{\cal{B}}_{\mu\nu}J^{\mu\nu}\right]
\end{align}  
and integrating out the pseudovector at tree level one ends up with the effective Lagrangian\footnote{Strictly speaking, integration of ${\cal{B}}$ brings corrections to the gauge kinetic terms of the form $g^2W_{\mu\nu}^aW^{\mu\nu a}$ and $g^{\prime 2}B_{\mu\nu}B^{\mu\nu}$. Such terms can however be reabsorbed by appropriate redefinition of the gauge fields and couplings, such that the kinetic terms remain canonically normalized.}
\begin{align}
{\cal{L}}_{{\it{eff}}}&\!=\!\frac{1}{m_{\cal{B}}^2}J_{\mu\nu}^iJ_i^{\mu\nu}\!\!=\sum_j\!\left[c_j^{(2)}{\cal{O}}_j^{(2)}\!+c_j^{(3)}{\cal{O}}_j^{(3)}\!+c_j^{(4)}{\cal{O}}_j^{(4)}\right]
\end{align}
where
\begin{align}
J_{\mu\nu}^{j}&\equiv t^jJ_{\mu\nu}=f_1~{\mathrm{tr}}\!\left(t^jf_{+\mu\nu}\right)+if_2~{\mathrm{tr}}\left(t^j[L_{\mu},L_{\nu}]\right)\nonumber\\
&\!\!\!\!\!\!\!\!\!\!\!\!\!\!+h_t({\bar{\psi}}_L\sigma_{\mu\nu}ut^juP_+\psi_R)+h_b({\bar{\psi}}_L\sigma_{\mu\nu}ut^juP_-\psi_R)+{\mathrm{h.c.}}
\end{align}
and use has to be made of the $SU(2)$ relation
\begin{align}
t_{ij}^at^a_{kl}&=\frac{1}{2}\delta_{il}\delta_{jk}-\frac{1}{4}\delta_{ij}\delta_{kl}
\end{align}
\begin{figure}[t]
\begin{center}
\includegraphics[width=2.6cm]{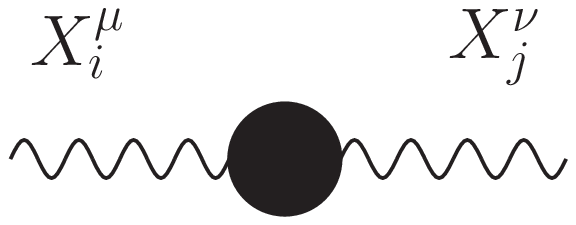}
\hspace{0.2in}
\includegraphics[width=3.0cm]{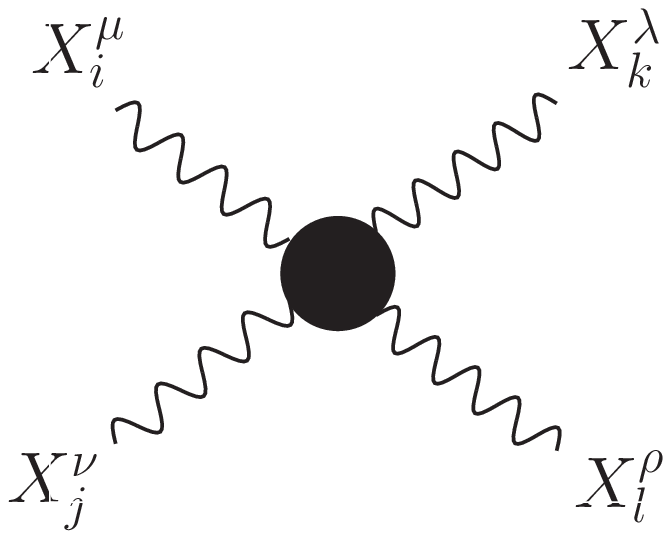}\\
\vspace{0.2in}
\includegraphics[width=3.0cm]{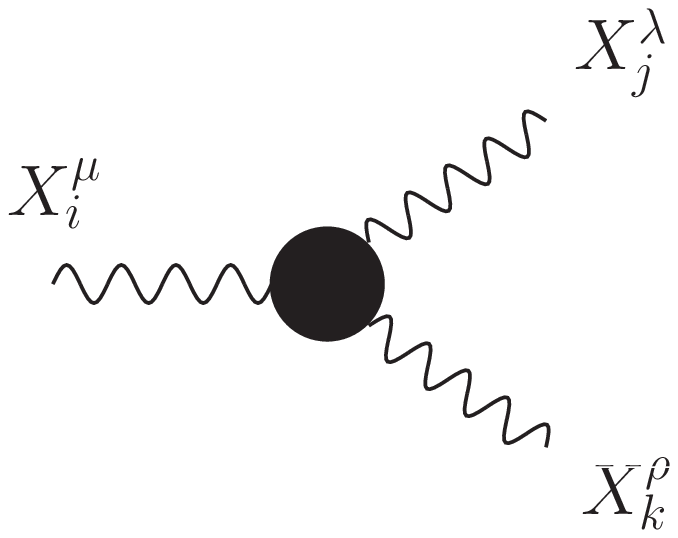}
\hspace{0.2in}
\includegraphics[width=2.7cm]{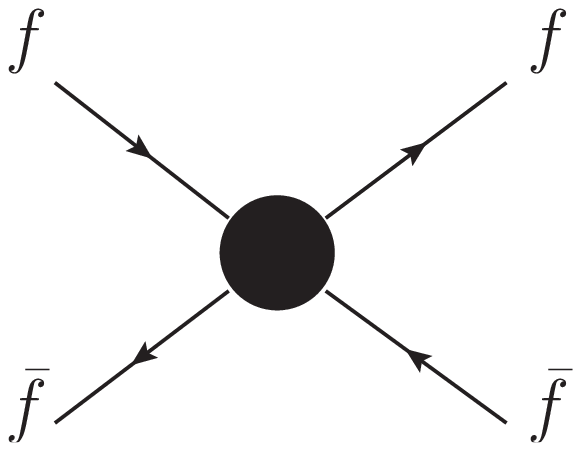}\\
\vspace{0.2in}
\includegraphics[width=2.7cm]{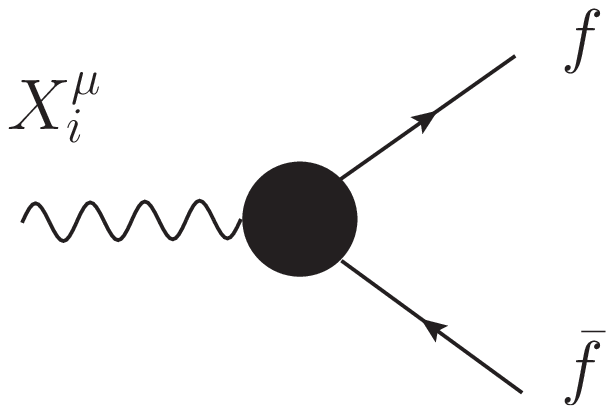}
\hspace{0.2in}
\includegraphics[width=2.9cm]{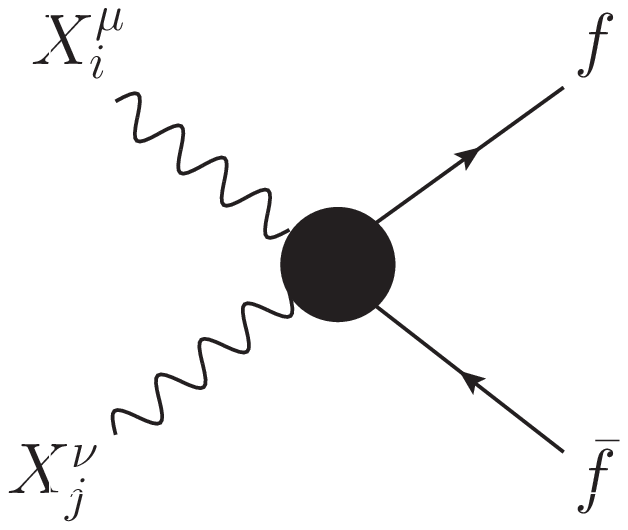}
\end{center}
\caption{\small{\it{Different low-energy topologies affected by pseudovector exchange. They correspond to anomalous Standard Model vertices, except the last topology, which is not present in the Standard Model.}}}\label{fig1}
\end{figure}
The effective operators ${\cal{O}}_j^{(i)}$ describe the anomalous oblique, triple and quartic vertices depicted in Fig.~\ref{fig1}. They are listed, together with their corresponding coefficients, in Table~\ref{tab1}. From it one would naively conclude that two, three and four-point vertices get modified. However, if one computes explicitly the corresponding diagrams with pseudovector exchange, one realizes that only for the four-point vertices there is actual resonance propagation. Pseudovector propagation in two and three-point vertices is forbidden by $P$ and $C$ conservation, and the contribution reported in Table~\ref{tab1} corresponds to a contact term interaction. 

Such contact interactions are licit and physically relevant provided that they do not spoil the consistency of the theory at high energies. Even though I have not committed to a UV completion of the present model, any strongly-coupled scenario implicitly requires duality, above the deconfinement scale, to a theory of more fundamental constituents. The paradigmatic example is QCD, where hadrons are made of quarks and gluons. 

In order to link the confined and deconfined phases, it will be convenient to adopt the language of dispersion relations. If the contact terms do not conform to the expected high-energy behavior of the theory, then the model has to be extended with the addition of local operators such that the mismatch is avoided. A similar strategy was adopted in QCD to assess the impact of the $b_1(1235)$ meson in chiral dynamics~\cite{Ecker:2007us}.  

Consider the two-point function $\Pi_{WB}$, defined as
\begin{align}
\Pi_{WB}^{\mu\nu}(q)&=-\frac{\delta^2}{\delta W_{\mu}^3\delta B_{\nu}}{\cal{L}}=\left(\frac{q^{\mu}q^{\nu}}{q^2}-g^{\mu\nu}\right)\Pi_{WB}(q^2)
\end{align}
This correlator is an order parameter of custodial symmetry breaking. At high energies, where the global group $SU(2)_L\times SU(2)_R$ is unbroken, one expects perturbative contributions in terms of the fundamental constituents to cancel to all orders. Nonzero contributions should come from nontrivial vacuum condensates in the operator power expansion. The correlator is therefore power-suppressed at high energies and expected to satisfy an unsubtracted dispersion relation
\begin{align}\label{WB}
\Pi_{WB}(q^2)&=\int_0^{\infty}\frac{dt}{t-q^2}\frac{1}{\pi}{\mathrm{Im}}\Pi_{WB}(t)
\end{align}
\begin{table}[t]
\begin{center}
\begin{tabular}{lr}
\hline
Operators ${\cal{O}}_j^{(i)}$ &\,\,\,\, Coefficients $c_j^{(i)}$\\[0.1cm]
\hline
${\cal{O}}_1^{(2)}=gg^{\prime}\langle W_{\mu\nu}\tau_L\rangle B^{\mu\nu}$ &\,\,\,\, $f_1^2/m_{\cal{B}}^2$ \\[0.1cm]
\hline
${\cal{O}}_1^{(3)}=g\langle W_{\mu\nu}[L^{\mu},L^{\nu}]\rangle$ &\,\,\,\, $if_1f_2/m_{\cal{B}}^2$ \\
${\cal{O}}_2^{(3)}=g^{\prime}B_{\mu\nu}\langle \tau_L[L^{\mu},L^{\nu}]\rangle$ &\,\,\,\, $if_1f_2/m_{\cal{B}}^2$ \\
${\cal{O}}_3^{(3)}=g{\bar{\psi}}_L\sigma_{\mu\nu}W^{\mu\nu}UP_+\psi_R$ &\,\,\,\, $f_1h_t/m_{\cal{B}}^2$\\
${\cal{O}}_4^{(3)}=g{\bar{\psi}}_L\sigma_{\mu\nu}W^{\mu\nu}UP_-\psi_R$ &\,\,\,\, $f_1h_b/m_{\cal{B}}^2$ \\
${\cal{O}}_5^{(3)}=g^{\prime}{\bar{\psi}}_L\sigma_{\mu\nu}B^{\mu\nu}Ut_3P_+\psi_R$ &\,\,\,\, $f_1h_t/m_{\cal{B}}^2$ \\
${\cal{O}}_6^{(3)}=g^{\prime}{\bar{\psi}}_L\sigma_{\mu\nu}B^{\mu\nu}Ut_3P_-\psi_R$ &\,\,\,\, $f_1h_b/m_{\cal{B}}^2$ \\[0.1cm]
\hline
${\cal{O}}_1^{(4)}=\langle L_{\mu}L_{\nu}\rangle\langle L^{\mu}L^{\nu}\rangle$ &\,\,\,\, $-f_2^2/m_{\cal{B}}^2$ \\
${\cal{O}}_2^{(4)}=\langle L_{\mu}L^{\mu}\rangle\langle L_{\nu}L^{\nu}\rangle$ &\,\,\,\, $f_2^2/m_{\cal{B}}^2$ \\
${\cal{O}}_3^{(4)}=({\bar{\psi}}_L\sigma_{\mu\nu}UP_+\psi_R)({\bar{\psi}}_L\sigma_{\mu\nu}UP_+\psi_R)$ &\,\,\,\, $h_t^2/(4m_{\cal{B}}^2)$ \\
${\cal{O}}_4^{(4)}=({\bar{\psi}}_L\sigma_{\mu\nu}UP_-\psi_R)({\bar{\psi}}_L\sigma_{\mu\nu}UP_-\psi_R)$ &\,\,\,\, $h_b^2/(4m_{\cal{B}}^2)$ \\
${\cal{O}}_5^{(4)}=({\bar{\psi}}_L\sigma_{\mu\nu}UP_-\psi_R)({\bar{\psi}}_L\sigma_{\mu\nu}UP_+\psi_R)$ &\,\,\,\, $h_th_b/(2m_{\cal{B}}^2)$ \\
${\cal{O}}_6^{(4)}=({\bar{\psi}}_L\sigma_{\mu\nu}UP_-\psi_R)({\bar{\psi}}_RP_+\sigma_{\mu\nu}U^{\dagger}\psi_L)$ &\,\,\,\, $-h_bh_t^*/(2m_{\cal{B}}^2)$ \\
${\cal{O}}_7^{(4)}=({\bar{\psi}}_L\sigma_{\mu\nu}UP_-\psi_R)({\bar{\psi}}_RP_-\sigma_{\mu\nu}U^{\dagger}\psi_L)$ &\,\,\,\, $-h_bh_b^*/(2m_{\cal{B}}^2)$ \\
${\cal{O}}_8^{(4)}=({\bar{\psi}}_L\sigma_{\mu\nu}UP_+\psi_R)({\bar{\psi}}_RP_+\sigma_{\mu\nu}U^{\dagger}\psi_L)$ &\,\,\,\, $-h_th_t^*/(2m_{\cal{B}}^2)$ \\
${\cal{O}}_9^{(4)}={\bar{\psi}}_L\sigma_{\mu\nu}[L^{\mu},L^{\nu}]UP_+\psi_R$ &\,\,\,\, $if_2h_t/m_{\cal{B}}^2$ \\
${\cal{O}}_{10}^{(4)}={\bar{\psi}}_L\sigma_{\mu\nu}[L^{\mu},L^{\nu}]UP_-\psi_R$ &\,\,\,\, $if_2h_b/m_{\cal{B}}^2$ \\[0.1cm]
\hline
\end{tabular}
\end{center}
\caption{\small{\it{Low-energy effective operators and coefficients resulting from pseudovector integration. Complex conjugate operators are implicitly understood. As discussed in the main text, internal dynamical consistency requires the presence of additional counterterms. As a result, there is only net contribution from the four-fermion operators.}}}\label{tab1}
\end{table}

The contribution from the pseudovector to the absorptive part above is exactly zero, because it is not a propagating mode. Actually, $\Pi_{WB}^{({\cal{B}})}(q^2)=-gg^{\prime}\frac{f_1^2}{m_{\cal{B}}^2}q^2$ thus effectively generating a subtraction, which would indicate breaking of custodial symmetry in the deep UV and is therefore ruled out by general principles. This means that the model introduced in Eq.~(\ref{model}) needs the addition of counterterms such that Eq.~(\ref{WB}) holds true. To be more precise, consistency is restored when the counterterms balance out the naive pseudovector contribution. Therefore, the pseudovector model is consistent only when there is no net effect on $\Pi_{WB}$. This in particular means that there is no pseudovector contribution to the S parameter. 

Three-point vertices can likewise be examined through a dispersive approach. For QCD-inspired UV completions, ${\mathrm{Im}}\Pi_{XX}$ scales like a constant in the deep UV. Under this assumption the three-point form factors of Fig.~\ref{fig1} must obey once-subtracted dispersion relations of the form
\begin{align}
F_{X\to {\bar{f}}f}(q^2)&=1+q^2\int_0^{\infty}\frac{dt}{t(t-q^2)}\frac{1}{\pi}{\mathrm{Im}}F_{X\to {\bar{f}}f}(t)
\end{align}      
and likewise for $X^i\to X^jX^k$. The subtraction above is fixed by electric charge normalization. Again, the fact that the pseudovector does not contribute to the absorptive part means that counterterms are needed to avoid the appearance of an unphysical subtraction. These counterterms precisely balance out the naive pseudovector contribution. 

Notice that the situation with the four-point functions is rather different. There pseudovector exchange has a nonzero absorptive part ($C$ and $P$ no longer prevent resonance propagation) and its contribution resembles the one of regular vector resonances. However, counterterms are also needed in this case. To see this consider the longitudinal gauge boson scattering $W_L^aW_L^b\to W_L^cW_L^d$. By the equivalence theorem, at high energies this corresponds to the scattering of Goldstones $\phi^a\phi^b\to \phi^c\phi^d$, up to corrections of ${\cal{O}}(m_W/\sqrt{s})$. Goldstone scattering is determined by a single function ${\cal{A}}(x,y,z)$, where $x,y,z$ are the Mandelstam variables. Since ${\cal{A}}$ is symmetric in the last two arguments, only the first argument need to be kept above and one can define ${\cal{A}}(x)\equiv {\cal{A}}(x,y,z)={\cal{A}}(x,z,y)$. The total amplitude reads
\begin{align}
{\cal{M}}(\phi^a\phi^b\to \phi^c\phi^d)&\!=\delta^{ab}\delta^{cd}\!{\cal{A}}(s)+\delta^{ac}\delta^{bd}\!{\cal{A}}(t)+\delta^{ad}\delta^{bc}\!{\cal{A}}(u)
\end{align}
In elastic channels with $s\leftrightarrow u$ symmetry, the Froissart bound~\cite{Froissart:1961ux} leads to the following forward (once-subtracted) dispersion relation  
\begin{align}\label{su}
A(\nu,t=0)&=c+\nu^2\!\int_0^{\infty}\!\!\frac{d\eta}{\eta(\eta^2-\nu^2)}\frac{1}{\pi}{\mathrm{Im}}A(\eta,0)
\end{align}
where $\nu=\frac{1}{2}(s-u)$. If the new strong dynamics obeys the Froissart bound at asymptotically high energies, then at high $\nu^2$ the previous amplitudes should go like a constant. Explicit evaluation of pseudovector exchange shows that pseudovector exchange is non-propagating in Goldstone scattering. This generates a quadratic growth in~(\ref{su}), which again calls for the addition of counterterms.

The non-propagating character of ${\cal{B}}$ in longitudinal high-energy scattering also suggests that the four-point vertices $X^aX^b\to {\bar{f}}f$ will likewise need counterterms. Actually, explicit calculation shows that pseudovector contributions to $\phi^a\phi^b\to {\bar{f}}f$ only generate a contact term. 

A consistent pseudovector model thus requires to be enlarged with the following counterterms:  
\begin{align}
{\cal{L}}_C&=-\frac{f_1^2}{m_{\cal{B}}^2}{\cal{O}}_{1}^{(2)}-i\frac{f_1f_2}{m_{\cal{B}}^2}\left({\cal{O}}_1^{(3)}+{\cal{O}}_2^{(3)}\right)\nonumber\\
&-\frac{f_1h_t}{m_{\cal{B}}^2}\left({\cal{O}}_3^{(3)}+{\cal{O}}_5^{(3)}\right)-\frac{f_1h_b}{m_{\cal{B}}^2}\left({\cal{O}}_4^{(3)}+{\cal{O}}_6^{(3)}\right)+{\mathrm{h.c.}}\nonumber\\
&+\frac{f_2^2}{m_{\cal{B}}^2}\left({\cal{O}}_1^{(4)}-{\cal{O}}_2^{(4)}\right)-i\frac{f_2h_t}{m_{\cal{B}}^2}{\cal{O}}_9^{(4)}-i\frac{f_2h_b}{m_{\cal{B}}^2}{\cal{O}}_{10}^{(4)}+{\mathrm{h.c.}}
\end{align}
which precisely balance out the naive pseudovector low-energy contributions. The net result is that, with the exception of four-fermion topologies, low-energy traces of the pseudovector are obliterated. 

Since the conclusions above are based on generic high-energy properties, they are independent of particular resonance models above the TeV scale. However, in order to understand the physical entity of the local terms introduced above, consider enlarging the model we have presented here to a model of large-$N_c$ by inserting a full tower of pseudovector excitations all the way up to infinity. In this case, counterterms would be absent and consistency with dispersion relations would instead be satisfied through the existence of a set of sum rules, where the added contributions of the states in the tower would be required to non-trivially vanish. Unlike more conventional sum rules, those would however not be spectral sum rules. The counterterms introduced above can therefore be understood as the integrated-out contribution of a full tower of pseudovectors, with the exception of its lowest-lying mode, which at the energies we are interested in is light enough to be dynamical. 


\section{Direct detection at colliders}\label{sec:V}

The results of the previous Section show how elusive the light pseudovector can be for indirect detection. Its low-energy traces reduce to anomalous four-fermion interactions, which turn out to be mass-suppressed. Therefore, only top quark four-point vertices are nonnegligible with $c_{4t}\sim {\cal{O}}(1/\Lambda^2)$, yet extremely challenging to probe. Unlike most of the existing new-physics candidates, a pseudovector can in practice only be detected via direct searches.

In this Section I will discuss the most promising modes for direct detection of pseudovectors in a rather qualitative way. This Section is intended to provide a broad-brush picture of the predominant channels for discovery. A more detailed and quantitative study of the specific phenomenology and prospects for detection is postponed to a future publication.     

Inspection of Eq.~(\ref{modelB}) shows that the neutral pseudovector component decays predominantly into a gauge boson pair and a top pair, ${\cal{B}}^{0}\to W^+W^-$ and ${\cal{B}}^0\to {\bar{t}}t$. The charged component instead is dominated by ${\cal{B}}^{\pm}\to W^{\pm}Z$ and ${\cal{B}}^{\pm}\to {\bar{t}}b$. Of course, if the charged component is heavy enough, the radiative channels ${\cal{B}}^{\pm}\to {\cal{B}}^0 W^{\pm}$ will also be open. Explicit computation gives  
\begin{align}\label{decay}
\Gamma_{\cal{B}}^{(W^+W^-)}&=\frac{g^4}{48\pi}\left(\frac{f_2^2}{m_{\cal{B}}}\right)\frac{(1+2x_W^2)\sqrt{1-4x_W^2}}{x_W^2}
\end{align}
where $x_W=\displaystyle\frac{m_W}{m_{\cal{B}}}$. Incidentally, if one compares~(\ref{decay}) with the same decay for a regular vector~\cite{Cata:2009iy}, one finds that the pseudovector has a relative ${\cal{O}}(v^2/m_{\cal{B}}^2)$ suppression. A pseudovector is therefore generically narrower than a regular vector. 

Assuming CP-conserving interactions, the decay into a top pair is given by 
\begin{align}
\Gamma_{\cal{B}}^{({\bar{t}}t)}&=\frac{h_t^2}{24\pi}m_{\cal{B}}(1-4x_t^2)^{3/2}
\end{align}
where $x_t=\displaystyle\frac{m_t}{m_{\cal{B}}}$. Notice that since $\Gamma_{\cal{B}}^{(WW)}\!\!\sim{\cal{O}}(g^2m_{\cal{B}})$ and $\Gamma_{\cal{B}}^{({\bar{t}}t)}\!\!\sim{\cal{O}}(h_t^2m_{\cal{B}})$, both decay modes are of comparable size:  
\begin{align}
\frac{\Gamma_{\cal{B}}^{(W^+W^-)}}{\Gamma_{\cal{B}}^{({\bar{f}}f)}}\sim \frac{g^2}{y_t^2}\sim{\cal{O}}(1)
\end{align}
Numerically, for the mass-range of interest, $\Gamma_{\cal{B}}\sim 2\%\, m_{\cal{B}}$.

\begin{figure}[t]
\begin{center}
\includegraphics[width=3.2cm]{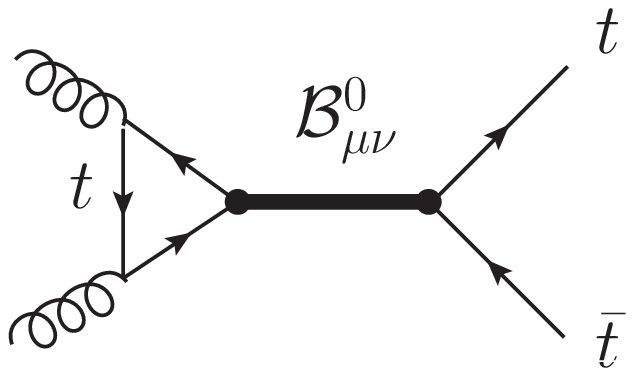}
\hspace{0.2in}
\includegraphics[width=3.4cm]{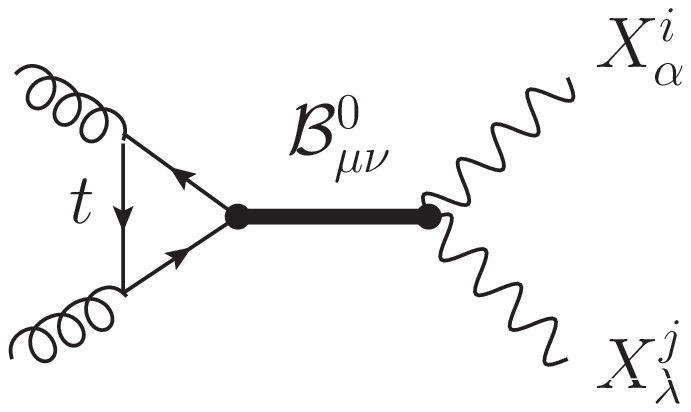}\\
\vspace{0.2in}
\includegraphics[width=3.8cm]{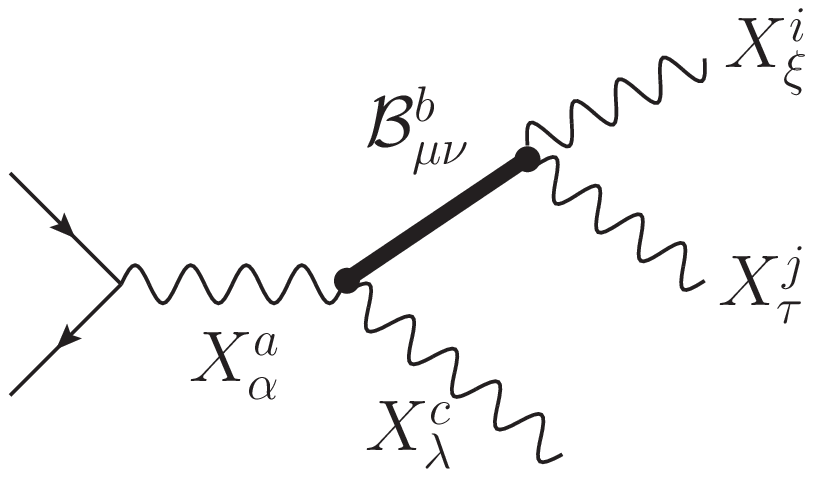}
\hspace{0.2in}
\includegraphics[width=3.8cm]{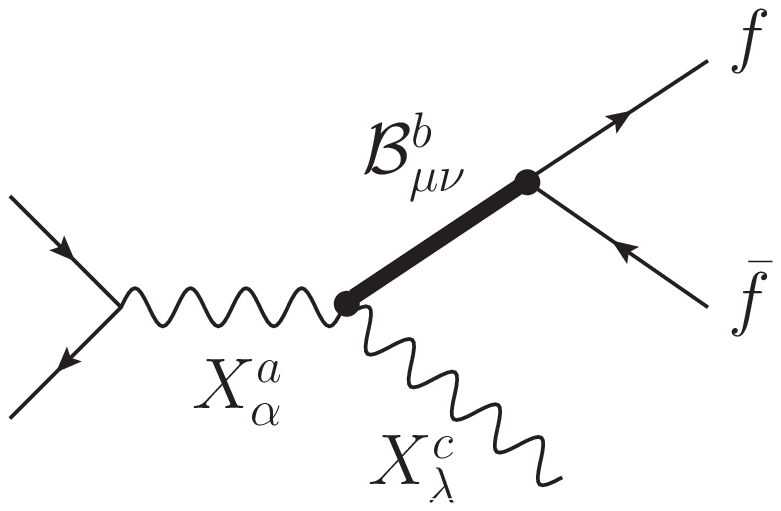}
\end{center}
\caption{\small{\it{Dominant modes for detection at colliders. See main text for a detailed discussion.}}}\label{fig:2}
\end{figure}
Regarding its detection at the LHC, the pseudovector is most favorably produced through gluon fusion and subsequent decay into ${\bar{t}}t$ jets and $W^+W^-$ (see upper panel in Fig.~\ref{fig:2}). The latter decay mode offers the cleaner signal for detection, and direct searches should not differ much from the ones already performed for the Higgs decay $h\to W^+W^-$, where both $W$ decay leptonically, and in searches for heavier scalar states~\cite{ATLAS}. However, due to the high pseudovector mass, it might be more efficient to let one of the W decay into a quark pair. The main background $W+\,$jets can then be efficiently rejected by applying a cut on the neutrino transverse momentum~\cite{Dittmaier:2012vm}. Unlike the Higgs, a specific feature of ${\cal{B}}$ is that ${\cal{B}}^0\to ZZ$ is also loop-induced. Therefore, there should be an excess of $WW$ production over $ZZ$. 

Notice also that, contrary to the vector scenario studied in~\cite{Cata:2009iy}, Drell-Yan production does not occur in this case, since a (propagating) pseudovector cannot couple to a single gauge boson. An interesting alternative is to consider the pseudovector in associated production with a gauge boson, as depicted in the lower panel of Fig.~\ref{fig:2}. The possible decay modes are $W{\bar{t}}t$, $Z{\bar{t}}b$, $WWW$ and $WZZ$. Among them, the cleanest ones are $Z{\bar{t}}b$ with leptonic decay of the $Z$, and $WZZ$ with one $Z$ decaying hadronically\footnote{As discussed in~\cite{Cata:2009iy}, the most efficient background rejection for triple gauge production takes place when only two of them decay leptonically, at least one of them being a $Z$.}. The pseudovector should then show up by scanning the invariant mass distribution of the ${\bar{t}}b$ and $WZ$ pairs, respectively.   

At the ILC, detection through two-body decay is almost excluded, since it is suppressed by powers of the electron mass. Therefore, the only clean signature at the ILC is the associated production of ${\cal{B}}^{\pm}$ with $W^{\mp}$. The cleanest signatures are $W{\bar{t}}b$, with $W$ decaying into leptons, and $WWZ$, with one $W$ decaying into quarks. The pseudovector can be detected in the invariant mass distribution of the ${\bar{t}}b$ and $WZ$ pairs, respectively.   
  

\section{Conclusions}\label{sec:VI}

The most pressing issue in particle physics is to pin down, or at least shed some light on the dynamical mechanism that triggers electroweak symmetry breaking. The existence of a light scalar with properties close to the Standard Model Higgs does not resolve the issue but poses additional requirements that this underlying mechanism must fulfill, namely (i) it must provide a mechanism to stabilize the light scalar mass; and (ii) it should manifest itself around the TeV scale yet complying with the strongly-constrained deviations from the Standard Model paradigm.

Dynamical symmetry breaking is a long-standing candidate for such a mechanism. In such scenarios a light scalar can be easily accommodated as a pseudo-Goldstone boson of an underlying broken global symmetry. However, light states (below the TeV scale) are hard to reconcile with the constraints of electroweak precision data. The smallness of Standard Model deviations suggests that new physics states should leave a rather subtle imprint.
   
In this paper I have examined the viability of sub-TeV states when confronted with experimental constraints. In particular, I have considered a scenario of new dynamics invariant under parity and $SU(2)_L\times SU(2)_R$, spontaneously broken down to the custodial $SU(2)_V$. The lowest-lying resonance in the spectrum is a pseudovector, described as a Kalb-Ramond antisymmetric tensor field ${\cal{B}}_{\mu\nu}$ with couplings to the Standard Model fields. I have shown that such a state is rather elusive in the low-energy theory, not because of additional {\it{ad hoc}} suppressions of its couplings, but by the requirement that the strong dynamics possess a consistent asymptotically-free UV completion. Applying this criteria one can show that the pseudovector evades the constraints coming from oblique parameters, electric and magnetic dipole moments, as well as triple and quartic gauge-boson vertices. Indirect traces thereof would only affect third-family four-quark operators, which are presently poorly constrained.  

Consequently, the existence of a ${\cal{B}}_{\mu\nu}$ could only be tested by direct detection. At an hadronic collider like the LHC, it is mostly produced through gluon-fusion and subsequent decay into a $W^+W^-$ pair or a top dijet. In contrast, at the ILC it could only be detected in its associated production with gauge bosons. Its signal could be isolated in $W{\bar{t}}b$ and $WWZ$ by scanning through the invariant mass distribution of the ${\bar{t}}b$ and $WZ$ pairs, respectively. A more detailed quantitative analysis of the collider phenomenology would require a dedicated paper and is left for future work.

\section*{Acknowledgments}
I thank Gehard Buchalla for carefully reading the manuscript and providing useful comments and suggestions. This project was supported in part by the ERC Advanced Grant project 
'FLAVOUR' (267104) and the DFG cluster of excellence 'Origin and Structure of the Universe' 


\end{document}